\documentclass[a4paper,nobibnotes,nofootinbib]{revtex4}
\usepackage{amssymb}
\usepackage{amsmath}
\usepackage{epsfig}
\newcommand{\be}{\begin{equation}}
\newcommand{\ee}{\end{equation}}
\newcommand{\bea}{\begin{eqnarray}}
\newcommand{\eea}{\end{eqnarray}}

\newcommand{\D}{\displaystyle}
\newcommand{\g}{\gamma}
\newcommand{\f}{\frac}

\newcommand{\bra}{\langle}
\newcommand{\ket}{\rangle}
\newcommand{\intc}[1]{{\int\frac{d#1}{2i\pi}}}
\newcommand\lr[1]{{\left({#1}\right)}}
\begin{document}
\title{A QCD dipole formalism for forward-gluon production}
\author{Cyrille Marquet}\email{marquet@spht.saclay.cea.fr}
\affiliation{Service de physique th{\'e}orique, CEA/Saclay,
  91191 Gif-sur-Yvette cedex, France\footnote{%
URA 2306, unit{\'e} de recherche associ{\'e}e au CNRS.}}
\begin{abstract}

We derive inclusive and diffractive forward-gluon production in the scattering 
of a $q\bar q$ dipole off an arbitrary target in the high-energy eikonal 
approximation, suitable to study the saturation regime. We show how the 
inclusive cross-section is related to the total cross-section for the scattering 
of a colorless pair of gluons on the target: the gluon-production cross-section 
can be expressed as a convolution between this $gg$ dipole total cross-section 
and a dipole distribution. We then consider as an application the forward-jet 
production from an incident hadron and describe forward-jet production at HERA 
and Mueller-Navelet jets at Tevatron or LHC. We show how these measurements are 
related to the $q\bar q-gg$ or $gg-gg$ dipole-dipole cross-sections and why they 
are therefore well-suited for studying high-energy scattering in QCD.

\end{abstract}
\maketitle
\section{Introduction}
\label{1}

Understanding high-energy scattering in QCD has been the purpose of a lot 
of work over the past ten years. At high energies, as one approaches the 
unitarity limit, high-density phases of partons are created and non-linear 
effects become important. In this regime called saturation 
\cite{glr,glr+,mv,jimwlk,bk,bk+,golec}, the basic properties of perturbative 
QCD, such as factorization and linear evolution, breakdown. To study scattering 
near the unitarity limit, the $q\bar q$ dipole model \cite{dipole,mueller} has 
been developed. This formalism constructs the light-cone wavefunction of a 
dipole (a quark-antiquark pair in the color singlet state) in the leading 
logarithmic approximation. The size of the dipole provides a scale which is 
supposed to be much smaller than $1/\Lambda_{QCD}$ to justify the use of 
perturbation theory. As the energy increases, the dipole evolves and the 
wavefunction of this evolved dipole is described as a system of elementary 
dipoles. This formalism is well-suited to study high-energy scattering because 
when this system of dipoles scatters on a target, density effects and 
non-linearities that lead to saturation and unitarization of the scattering 
amplitude can be taken into account.

On a phenomenological point of view, the key point is to relate the experimental 
probes to the scattering of a dipole. In the case of lepton-hadron collisions 
this is straightforward: a lepton undergoes hadronic interactions via a virtual 
photon and one can view the interation as the splitting of the virtual photon 
into a $q\bar q$ pair which then interacts. This dipole picture has had great 
succes for the phenomenology of hard processes initiated by virtual photons. In 
hadron-hadron collisions however, there are no such virtual-photon probes. The 
probes are the final-state particules that we measure, {\it e.g.} a final-state 
gluon that we detect as a jet. Establishing a link between the scattering of a 
dipole and observables like jet cross-sections is of great theoretical and 
phenomenological interest in the prospect of the LHC, if one considers the 
impact that the dipole picture has had on HERA phenomenology for example and 
how the HERA measurements have helped understanding high-energy scattering.

First steps in this direction were taken in \cite{kope} where gluon radiation 
from a quark was calculated. In \cite{kovtuch}, Kovchegov and Tuchin derived the 
cross-section for inclusive gluon production in deep inelastic scaterring off a 
large nucleus. In both cases, the cross-sections are expressed in terms of the 
scattering of a $gg$ dipole (a gluon-gluon pair in the color singlet state) on 
the target. It is crucial in \cite{kovtuch} that the target is a nucleus since 
they work in an approximation in which all the scatterings on the nucleons 
happen via one or two-gluon exchanges \cite{muelkov}. On a phenomenological 
side, descriptions of Mueller-Navelet jets in \cite{marpes} and forward jets in 
\cite{mpr} are performed in terms of $q\bar q$ dipole-$q\bar q$ dipole 
scattering, leading to interesting results and predictions. The key point in 
these analysis is the description of the jet vertex in terms of $q\bar q$ 
dipoles: it is done using the equivalence between the $k_T-$factorization and 
the dipole factorization. The factorization formula used is therefore not 
justified when the dipole-dipole cross-section is not $k_T-$factorizable which 
could be the case at high energies.

In this paper, we consider the collision between a $q\bar q$ dipole and an 
arbitrary target and calculate the cross-section for the production of a gluon 
in the forward direction of the dipole. We work in the eikonal approximation 
valid for high-energy scattering in which the interaction with the target is 
described by Wilson lines. We derive formulae for the inclusive and diffractive 
cross-sections. We express the inclusive cross-section in terms of the 
scattering of a colorless gluon-gluon pair on the target. We generalize the 
result of \cite{kovtuch}, obtaining it in a very different framework and for the 
scattering off any target. We write the inclusive cross-section in the 
factorized form given below in equation (\ref{dipfact}): a convolution between 
an effective dipole distribution and the total cross-section for the scattering 
of a $gg$ dipole on the target; such a factorization does not happen for the 
diffractive cross-section. We obtain a very general formula that includes all 
numbers of gluon exchanges and non-linear quantum evolution.

We then take another step towards the description of observables in hadronic 
collisions as we extend our results to the case of an incident hadron instead of 
an incident $q\bar q$ pair. This is done by including linear evolution before 
the emission of the measured gluon and by using collinear factorization. We thus 
describe forward-jet cross-sections at hadronic colliders in terms of the 
scattering of a $gg$ dipole on the target, see later formula (\ref{fjint}) which 
exhibits a dipole factorization for the forward-jet probe and generalizes the 
formulation of \cite{marpes,mpr}.

Finally, as an application to our formulae, we consider the target to be a 
virtual photon, or equivalently a $q\bar q$ dipole, and obtain the forward-jet 
cross-section at HERA \cite{dis}. Its high-energy behavior is driven by the 
quantum evolution of the $gg$ dipole-$q\bar q$ dipole scattering, making the 
forward-jet measurement of particular interest to investigate high-energy 
scattering and unitarization in QCD. The same conclusion holds for the 
Mueller-Navelet jet cross-section \cite{mnj} at Tevatron or LHC that we also 
derive.

The plan of the paper is as follows. In Section 2, single inclusive 
forward-gluon production in the scattering of a $q\bar q$ dipole off an 
arbitrary target is calculated. The diffractive cross-section is derived in 
Section 3. The relation between the inclusive cross-section and the $gg$ 
dipole-target scattering is derived in Section 4. In section 5, extending the 
previous result to the case of an incident hadron and specifying different 
targets, we derive the Mueller-Navelet jet and forward-jet cross-sections. 
Section 6 concludes.

\section{Inclusive gluon production}
\label{2}

\begin{figure}[ht]
\begin{center}
\epsfig{file=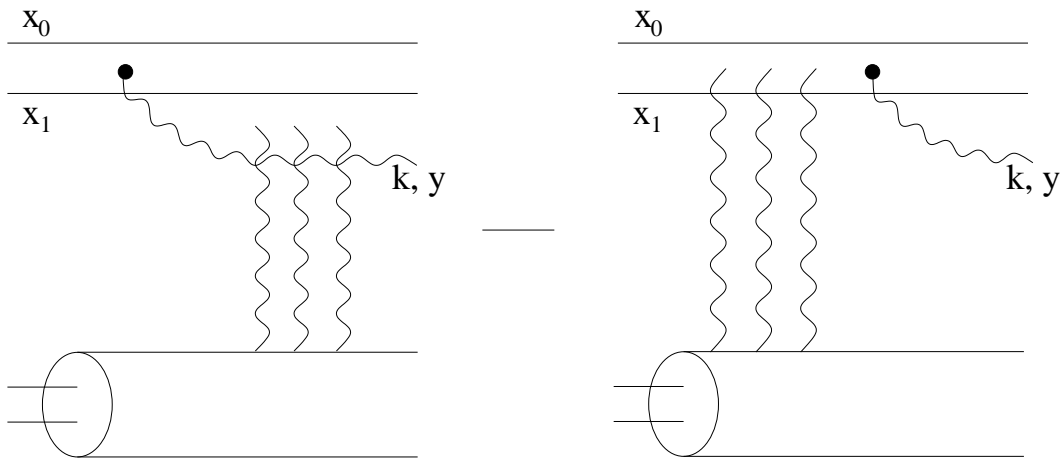,width=14cm}
\caption{Inclusive gluon production off a $q\bar q$ dipole. $x_0$ and $x_1:$ 
transverse 
coordinates 
of the quark and antiquark; $k$ and $y:$ transverse momentum and rapidity of the 
measured gluon. The black points represent emission of the gluon by either the 
quark or the antiquark. The vertical wavy lines represents the interaction with 
the target. The gluon is emitted before the interaction or after the interaction 
in which case it comes with a minus sign as explained in the text.}
\end{center}
\label{F1}
\end{figure}

In this section we derive single inclusive gluon production in the high-energy 
scattering of a $q\bar q$ dipole off an arbitrary target. We shall use 
light-cone coordinates with the incoming dipole being a right mover and work in 
the light-cone gauge $A_+=0.$ In this case, when the dipole passes through the 
target and interacts with its gauge fields, the dominant couplings are eikonal: 
the partonic components of the dipole (a quark, an antiquark and soft gluons) 
have frozen transverse coordinates and the gluon fields of the target do not 
vary during the interaction. This is justified since the incident dipole 
propagate at nearly the speed of light and its time of propagation through the 
target is shorter than the natural time scale on which the target fields vary.
The effect of the interaction with the target is that the components of the
dipole wavefunction pick up eikonal phases.

Let us be more specific. In Fig.1 is represented the production of a gluon of 
transverse momentum $k$ and rapidity $y$ off a dipole with transverse 
coordinates $x_0$ for the quark and $x_1$ for the antiquark. The size of the 
dipole $|x_0\!-\!x_1|$ is supposed to be small in order to justify the use of 
perturbation theory ($|x_0\!-\!x_1|\!\ll\! 1/\Lambda_{QCD}$). We work in a frame
in which the dipole rapidity is not too large so that the radiation of extra 
softer gluons is to be described by quantum evolution of the target. The 
necessity of the last term can be seen by considering the case where there are 
no interactions: the emission-after-interaction term is necessary to get zero 
gluon production without interaction.

The incident hadronic state is a colorless $q\bar q$ pair and has the following 
decomposition on the Fock states:
\be
|d\ket=|d\ket_0+|dg\ket_0\ .\label{decomp}\ee
The bare dipole $|d\ket_0$ is characterised by the wavefunction
\be
|d\ket_0=\sum_{\alpha\bar\alpha}
\f{\delta_{\alpha,\bar{\alpha}}}{\sqrt{N_c}}
|(\alpha,x_0),(\bar{\alpha},x_1)\ket \ee
where $\alpha$ and $\bar\alpha$ denote the colors of the quark and antiquark
respectively and where the 
transverse positions of the partons have been specified.
The $q\bar qg$ part of the dressed dipole $|dg\ket_0$ is characterised by the 
wavefunction
\be
|dg\ket_0=\f1{\sqrt{N_c}}\sum_{\alpha\bar\alpha\lambda a}\int d^2z dy\
\f{ig_s}{\pi}\left[\f{\epsilon_\lambda.(x_0\!-\!z)}{|x_0\!-\!z|^2}
-\f{\epsilon_\lambda.(x_1\!-\!z)}{|x_1\!-\!z|^2}
\right] 
T_{\bar\alpha \alpha}^a
|(\alpha,x_0),(\bar\alpha,x_1),(a,\lambda,z,y)\ket\ ,\label{softg}\ee
where the gluon is characterized by its color $a$, its polarization 
$\lambda$,
its rapidity $y$ and its transverse coordinate $z.$ $\epsilon_\lambda$ is the
transverse component of the gluon polarization vector and $T^a$ is a generator
of the fundamental representation of $SU(N_c).$ The term in brackets in 
(\ref{softg}) is the well-known wavefunction for the emission of a gluon off a 
$q\bar q$ dipole \cite{mueller}; the two contributions correspond to 
emission by the quark and antiquark respectively. The only assumption made to 
write down (\ref{softg}) is that the gluon is soft, 
that is its longitudinal fraction of momentum with respect to the incident 
dipole is small. As already mentionned, we work in a frame in which only bare or 
one-gluon components need to be considered in the wavefunction $|d\ket$; softer 
gluons will be included through the quantum evolution of the target. One then 
truncates the perturbative expansion (\ref{decomp}) at first order in the strong 
coupling constant $g_s.$

Let us denote the initial state of the target $|t\ket.$ The outgoing state is 
obtained from the incoming state $|d\ket\!\otimes\!|t\ket$ by action of the 
${\cal S}-$matrix. In the eikonal approximation, ${\cal S}$ acts on quarks and 
gluons as (see for example \cite{coursmuel,kovwie,heb}):
\be
{\cal S}|(\alpha,x)\ket\!\otimes\!|t\ket=
\sum_{\alpha'}\left[W_F(x)\right]_{\alpha\alpha'}
|(\alpha',x)\ket\!\otimes\!|t\ket\ ,\hspace{1.5cm}
{\cal S}|(a,\lambda,z,y)\ket\!\otimes\!|t\ket=\sum_{b}W_A^{ab}(z)
|(b,\lambda,z,y)\ket\!\otimes\!|t\ket\ ,\ee
where the phase shifts due to the interaction are described by $W_F$ and $W_A,$ 
the eikonal Wilson lines in the fundamental and adjoint representations 
respectively, corresponding to propagating quarks and gluons. They are given by
\begin{equation}
W_{F,A}(x)={\cal P}\exp\{ig_s\int dz_+T_{F,A}^a{\cal A}_-^a(x,z_+)\}
\end{equation}
with ${\cal A}_-$ the gauge field of the target and $T_{F,A}^a$ the 
generators of 
$SU(N_c)$ in the fundamental ($F$) or adjoint ($A$)
representations. ${\cal P}$ denotes an ordering in $z_+.$

Therefore the state 
$|\Psi_{out}\ket\!=\!{\cal S}|d\ket\!\otimes\!|t\ket$ emerging from the eikonal 
interaction reads $|\Psi_{out}\ket\!=\!|\Psi_1\ket\!+\!|\Psi_2\ket$ with the 
componants given by
\be
|\Psi_1\ket=\f1{\sqrt{N_c}}\sum_{\alpha\bar\alpha}
\left[W_F^\dagger(x_1)W_F(x_0)\right]_{\bar{\alpha}\alpha}
|(\alpha,x_0),(\bar{\alpha},x_1)\ket\otimes|t\ket\ ,\label{psi1}\ee
\bea
|\Psi_2\ket=\f1{\sqrt{N_c}}\sum_{\alpha\bar\alpha\lambda b}\int d^2z dy\
\f{ig_s}{\pi}
\left[\f{\epsilon_\lambda.(x_0\!-\!z)}{|x_0\!-\!z|^2}
-\f{\epsilon_\lambda.(x_1\!-\!z)}{|x_1\!-\!z|^2}\right]
\left[W_F^\dagger(x_1)T^aW_F(x_0)\right]_{\bar{\alpha}\alpha}\nonumber\\
W_A^{ab}(z)|(\alpha,x_0),(\bar\alpha,x_1),(b,\lambda,z,y)\ket\otimes|t\ket\ 
.\label{psi2}\eea
 $|\Psi_2\ket$ 
represents the first contribution 
pictured in Fig.1 while the second contribution is hidden in $|\Psi_1\ket.$ To 
see that, we must express the Fock states which appear in $|\Psi_{out}\ket$ in 
terms of physical states. At first order in $g_s$ one has: 
\bea
|(\alpha,x_0),(\bar{\alpha},x_1)\ket=|(\alpha,x_0),(\bar{\alpha},x_1)\ket_{phys}
-\sum_{\mu\lambda a}\int d^2z dy\
\f{ig_s}{\pi}\left[\f{\epsilon_\lambda.(x_0\!-\!z)}{|x_0\!-\!z|^2}
T_{\alpha \mu}^a|(\mu,x_0),(\bar\alpha,x_1),(a,\lambda,z,y)\ket
\right.\nonumber\\\left.-\f{\epsilon_\lambda.(x_1\!-\!z)}{|x_1\!-\!z|^2}T_{\mu 
\bar\alpha}^a
|(\alpha,x_0),(\mu,x_1),(a,\lambda,z,y)\ket\right]\ ,\label{phys1}\eea
\be
|(\alpha,x_0),(\bar\alpha,x_1),(b,\lambda,z,y)\ket=
|(\alpha,x_0),(\bar\alpha,x_1),(b,\lambda,z,y)\ket_{phys}\ .\label{phys2}\ee
One immediately sees that the emission-after-interaction term arises with a 
minus sign. Dropping the $q\bar q$ part which does not conribute to gluon 
production, using (\ref{psi1}), (\ref{psi2}), (\ref{phys1}), and (\ref{phys2}),  
the outgoing state can be rewritten
\bea
|\Psi_{out}\ket=
\frac{1}{\sqrt{N_c}}\sum_{\alpha\bar\alpha\lambda b}\int dy\int d^2z\ 
\f{ig_s}{\pi}\left\{\f{\epsilon_\lambda.(z\!-\!x_0)}{|z\!-\!x_0|^2}
\lr{\left[ W_F^\dagger(x_1)T^aW_F(x_0)\right]_{\bar{\alpha}\alpha}
W_A^{ab}(z)-\left[W_F^\dagger(x_1)W_F(x_0)T^b\right]_{\bar{\alpha}\alpha}}
\right.\nonumber\\
\left.-\f{\epsilon_\lambda.(z\!-\!x_1)}{|z\!-\!x_1|^2}\lr{\left[ 
W_F^\dagger(x_1)T^aW_F(x_0)
\right]_{\bar{\alpha}\alpha}
W_A^{ab}(z)-\left[T^bW_F^\dagger(x_1)W_F(x_0)\right]_{\bar{\alpha}\alpha}}
\right\}|(\alpha,x_0),(\bar\alpha,x_1),(b,\lambda,z,y)\ket_{phys}\otimes|t\ket\ 
.\label{outg}\eea
The different contributions contained in this wavefunction have a 
straightforward physical meaning: the term containing 
$\epsilon_\lambda.(z\!-\!x_0)$ comes from the emission of the gluon by the quark 
while the one containing $\epsilon_\lambda.(z\!-\!x_1)$ comes from to the 
emission by the antiquark. Moreover for both these terms, the contribution with 
three Wilson lines corresponds to the interation happening after the gluon 
emission while the contribution with two Wilson lines corresponds to the 
interation happening before. 

From the outgoing state (\ref{outg}), the gluon-production cross-section reads
\be
\f{d\sigma}{d^2kdy}(x_{01})=\f1{2(2\pi)^3}\int d^2b
\sum_{\lambda=\pm}\sum_{c=1}^{N_c^2-1}
\bra\Psi_{out}|a_{c,\lambda}^\dagger(k,y)a_{c,\lambda}(k,y)|\Psi_{out}\ket
\label{csec}\ee
where $a_{c,\lambda}^\dagger(k,y)$ and $a_{c,\lambda}(k,y)$ are respectively 
the creation and annihilation operators of a gluon with color $c,$ polarization 
$\lambda,$ rapidity $y$ and transverse momentum $k.$ $x_{01}\!=\!x_0\!-\!x_1$ 
is the size of the incoming dipole and $b\!=\!(x_0\!+\!x_1)/2$ is the
impact parameter. Let us rewrite the cross-section (\ref{csec}) using operators 
$a_{c,\lambda}(z,y)$ in transverse coordinate space that act on 
$|\Psi_{out}\ket:$
\be
\begin{array}{lll}\D\f{d\sigma}{d^2kdy}(x_{01})
&\D=\f1{4\pi}\int d^2b\ \f{d^2z_1}{2\pi}\f{d^2z_2}{2\pi}\ e^{ik.(z_2\!-\!z_1)}
\sum_{\lambda,c}\bra\Psi_{out}|
a_{c,\lambda}^\dagger(z_2,y)a_{c,\lambda}(z_1,y)|\Psi_{out}\ket\\\\
&\D=\f{\alpha_s}{\pi^2N_c}\int d^2b\ \f{d^2z_1}{2\pi} \f{d^2z_2}{2\pi}\ 
e^{ik.(z_2\!-\!z_1)}\bra t| P(x_0,x_1,z_1,z_2)|t\ket\ 
.\end{array}\label{defp}\ee
$z_1$ and $z_2$ represent now the transverse coordinates of the measured gluon 
in the amplitude and the complex conjugate amplitude respectively. 
$P(x_0,x_1,z_1,z_2)$ can be easily calculated from (\ref{outg}) and its 
definition (\ref{defp}) using
\be\begin{array}{lll}\D 
a_{c,\lambda}(z,y)|(\alpha,x_0),(\bar\alpha,x_1),(c',\lambda',z',y')\ket
\!=\!\delta_{cc'}\delta_{\lambda\lambda'}\delta^{(2)}(z\!-\!z')
\delta(y\!-\!y')|(\alpha,x_0),(\bar\alpha,x_1)\ket\ , \\\\
\D \bra(\alpha',x_0),(\bar\alpha',x_1)|(\alpha,x_0),(\bar\alpha,x_1)\ket
=\delta_{\alpha\alpha'}\delta_{\bar\alpha\bar\alpha'}\ . \end{array}\ee
It is given by
\bea
P(x_0,x_1,z_1,z_2)=\mbox{Tr}\left[\left\{\f{x_0\!-\!z_2}{|x_0\!-\!z_2|^2}
\lr{\left[ W_F^\dagger(x_0)T^cW_F(x_1)\right]W_A^{*cb}(z_2)
-\left[T^bW_F^\dagger(x_0)W_F(x_1)\right]}
\right.\right.\nonumber\\
\left.
-\f{x_1\!-\!z_2}{|x_1\!-\!z_2|^2}\lr{\left[
W_F^\dagger(x_0)T^cW_F(x_1)\right]W_A^{*cb}(z_2)
-\left[W_F^\dagger(x_0)W_F(x_1)T^b\right]}\right\}
\nonumber\\
.\left\{\f{x_0\!-\!z_1}{|x_0\!-\!z_1|^2}
\lr{\left[ W_F^\dagger(x_1)T^aW_F(x_0)\right]W_A^{ab}(z_1)
-\left[W_F^\dagger(x_1)W_F(x_0)T^b\right]}
\right.\nonumber\\
\left.\left.
-\f{x_1\!-\!z_1}{|x_1\!-\!z_1|^2}\lr{\left[
W_F^\dagger(x_1)T^aW_F(x_0)\right]W_A^{ab}(z_1)
-\left[T^bW_F^\dagger(x_1)W_F(x_0)\right]}\right\}\right].\label{traces}\eea
This expression contains terms with four, five, or six Wilson lines, however it 
can be reduced quite easily using the following formulae:
\be
\mbox{Tr}\lr{W_F^\dagger(x)T^aW_F(x)T^b}W_A^{ab}(z)=
\f12\mbox{Tr}\lr{W_A^\dagger(x)W_A(z)},\label{fierz1}\ee
\be
\mbox{Tr}\lr{W_F(x_1)T^aW_F^\dagger(x_1)W_F(x_0)T^aW_F^\dagger(x_0)}=
\f12\mbox{Tr}\lr{W_A^\dagger(x_1)W_A(x_0)}\ .\label{fierz2}\ee
These are obtained using the Fierz identity that relates the Wilson lines $W_F$ 
and $W_A:$
\be
\left[W_F(x)\right]_{ij}[W_F^{\dagger}(x)]_{kl}=\frac{1}{N_c}\ 
\delta_{il}\delta_{jk}
+2W_A^{ab}(x)T^a_{il}T^b_{kj}\ .\ee
In formula (\ref{traces}), terms which {\it a priori} contain six Wilson lines 
reduce to traces of two adjoint Wilson lines as $W_F^\dagger(x)W_F(x)=1.$ For 
the same reason, terms which display five Wilson lines contain only three and 
can be reduced using (\ref{fierz1}). Finally, terms with four Wilson lines are 
either trivial and equal to $C_FN_c$ or can be reduced using (\ref{fierz2}). At 
the end, the cross-section (\ref{defp}) obtained from $P(x_0,x_1,z_1,z_2)$ 
contains only traces of two adjoint Wilson lines and is given by
\bea
\f{d\sigma}{d^2kdy}(x_{01})=\f{\alpha_s}{2\pi^2N_c}
\int d^2b\int\f{d^2z_1}{2\pi}\f{d^2z_2}{2\pi}\ 
e^{ik.(z_2\!-\!z_1)}\sum_{i,j=0}^{1}(-1)^{i+j}
\f{(x_i\!-\!z_1).(x_j\!-\!z_2)}{|x_i\!-\!z_1|^2|x_j\!-\!z_2|^2}
\left\{\left\bra\mbox{Tr}\lr{W_A^\dagger(x_j)W_A(x_i)}\right\ket_t
\right.\nonumber\\\left.
-\left\bra\mbox{Tr}\lr{W_A^\dagger(x_j)W_A(z_1)}\right\ket_t
-\left\bra\mbox{Tr}\lr{W_A^\dagger(z_2)W_A(x_i)}\right\ket_t
+\left\bra\mbox{Tr}\lr{W_A^\dagger(z_2)W_A(z_1)}\right\ket_t\right\}
\label{final}\eea
where we have denoted $\bra t|\ .\ |t\ket\!=\!\bra\ .\ \ket_t.$

Interestingly enough, the only quantity involved in (\ref{final}) is the forward 
scattering amplitude of a colorless pair of gluons on the target:
\be 
T(x,x',Y\!-\!y)=
1-\f1{N_c^2\!-\!1}\left\bra\mbox{Tr}\lr{W_A^\dagger(x')W_A(x)}\right\ket_t\ . 
\label{gdip}\ee
This scattering amplitude is present in (\ref{final}) for all the dipoles 
involved in the process. It contains the scatterings with all numbers of gluon 
exchanges and, via its quantum evolution, it also contains the emissions of 
gluons softer than the measured one $(k,y).$ Indeed, if $Y$ is the total 
rapidity, the amplitude (\ref{gdip}) depends on $Y\!-\!y.$ The cross-section 
(\ref{final}) is made of four contributions, each of them proportional to one of 
the $(x_i\!-\!z_1).(x_j\!-\!z_2)$ ($i,j=0,1$), for which the gluon is emitted 
from the (anti)quark at transverse coordinate $x_i$ in the amplitude and $x_j$ 
in the complex conjugate amplitude. Each contribution itself contains four 
amplitudes $T$ whose physical origins are the following: $T(x_i,x_j,Y\!-\!y)$ 
corresponds to the gluon being emitted after the interaction both in the 
amplitude and in the complex conjugate amplitude in which case it does not 
interact; 
$T(x_i,z_2,Y\!-\!y)$ corresponds to emissions after the interaction in the 
amplitude and before the interaction in the complex conjugate amplitude and 
vice-versa for $T(z_1,x_j,Y\!-\!y);$ finally $T(z_1,z_2,Y\!-\!y)$ corresponds to 
emissions  before the interaction both in the amplitude and in the complex 
conjugate amplitude.

The scattering amplitude $T(x,x',Y\!-\!y)$ is related to the total 
cross-section $\sigma_{(gg)t}(x\!-\!x',Y\!-\!y)$ for the scattering of a gluon 
dipole $(gg)$ of size $x\!-\!x'$ on the target, with rapidity $Y\!-\!y$:
\be
\sigma_{(gg)t}(x\!-\!x',Y\!-\!y)=2\int d^2\lr{\f{x\!+\!x'}2}
T(x,x',Y\!-\!y)\ .\label{csection}\ee
To compute this cross-section, one has to evaluate the target 
averaging $\bra\ .\ \ket_t$ contained in $T$ (see formula (\ref{gdip})) which 
amounts to calculating averages of Wilson lines in the target wavefunction. A 
lot of studies are devoted to this problem as we briefly discuss later in 
Section V. Here we only establish the link between the observable (\ref{final}) 
and the dipole amplitude (\ref{gdip}).

Formula (\ref{final}) generalizes the result of 
\cite{kovch,kovtuch} where the same cross-section has been derived for a target 
nucleus in the quasi-classical approximation of \cite{muelkov} in which the 
scatterings on each nucleon happen via one or two gluon exchanges. It is quite 
remarkable that only one dynamical quantity appears in the result and this is 
what will allow us to write the inclusive cross-section (\ref{final}) in a 
factorized form as is shown in Section 4. But first, we shall calculate the 
diffractive gluon-production cross-section to exhibit the difference with the 
inclusive one and see that in this case, more than one dynamical quantity is 
involved.

\section{Diffractive gluon production}
\label{3}

\begin{figure}[ht]
\begin{center}
\epsfig{file=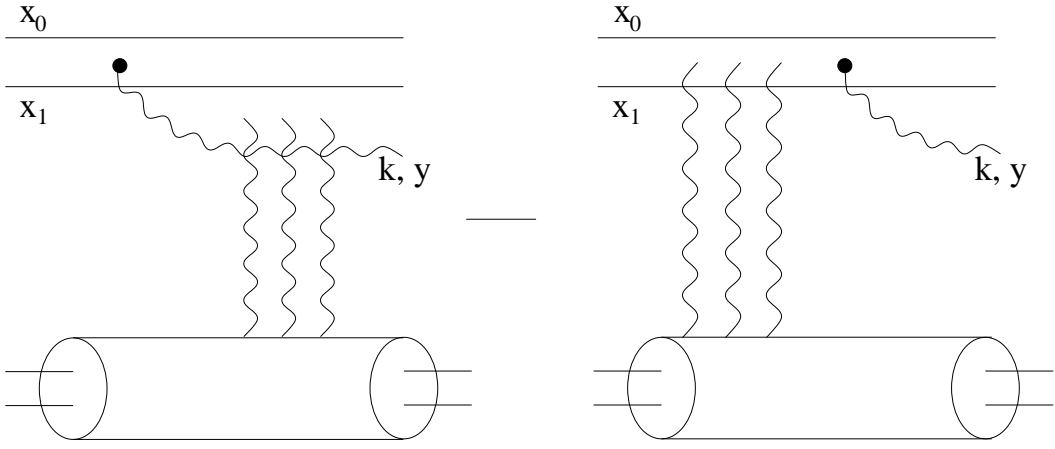,width=14cm}
\caption{Diffractive gluon production off a $q\bar q$ dipole. The object 
exchanged between the $q\bar q$ pair and the target is a color singlet and the 
target does not break up. Notations: see Fig.1.}
\end{center}
\label{F2}
\end{figure} 

The diffractive cross-section for gluon production in a $q\bar q$ dipole-target 
collision is pictured in Fig.2 where we define the diffractive process as one in 
which the outgoing wavefunction is in a color singlet state and in which the 
target does not break up. The cross-section is calculated in the same way than 
the inclusive one in the previous section. The difference is that we now have to 
project the outgoing state $|\Psi_{out}\ket$ on the subspace of color-singlet 
states. The final state containing diffractive gluon production is 
$|\Psi_{sing}\ket\!=\! P_{sing}|\Psi_{out}\ket$ where the projector on the 
color-singlet states $P_{sing}$ is given by:
\bea
P_{sing}=\f1{N_c}\int d^2x\ 
d^2x'\lr{\sum_{\mu,\nu}\delta_{\mu,\nu}|(\mu,x),(\nu,x')\ket}
\lr{\sum_{\alpha,\bar\alpha}\delta_{\alpha,\bar\alpha}
\bra(\alpha,x),(\bar\alpha,x')|}
+\f1{C_FN_c}\sum_{\lambda}\int d^2x\ d^2x'\ d^2zdy\nonumber\\
\lr{\sum_{\mu,\nu,b}T^b_{\nu\mu}|(\mu,x),(\nu,x'),(b,\lambda,z,y)\ket}
\lr{\sum_{\alpha,\bar\alpha,a}T^a_{\alpha \bar\alpha}
\bra(\alpha,x),(\bar\alpha,x'),(a,\lambda,z,y)|}\ \label{psing}.
\eea
One obtains with (\ref{psi1}), (\ref{psi2}), and (\ref{psing}):
\be
P_{sing}|\Psi_1\ket=\f1{N_c}\mbox{Tr}
\lr{W_F^\dagger(x_1)W_F(x_0)}|d\ket_0\otimes|t\ket\ ,\ee
\bea
P_{sing}|\Psi_2\ket=\f1{C_FN_c}\f1{\sqrt{N_c}}\sum_{\alpha\bar\alpha\lambda 
a}\int d^2z dy\ \f{ig_s}{\pi}
\left[\f{\epsilon_\lambda.(x_0\!-\!z)}{|x_0\!-\!z|^2}
-\f{\epsilon_\lambda.(x_1\!-\!z)}{|x_1\!-\!z|^2}\right]
\mbox{Tr}\lr{W_F^\dagger(x_1)T^bW_F(x_0)T^c}
W_A^{bc}(z)\nonumber\\T^a_{\bar\alpha 
\alpha}|(\alpha,x_0),(\bar\alpha,x_1),(a,\lambda,z,y)\ket\otimes|t\ket\ .\eea
$P_{sing}|\Psi_2\ket$ is the contribution of the emission-before-interaction 
term (first graph in Fig.2). Writing $|d\ket_0\!=\!|d\ket\!-\!|dg\ket_0$ in 
$P_{sing}|\Psi_1\ket$ makes the emission-after-interaction term arise (second 
graph in Fig.2). One is then able to express $|\Psi_{sing}\ket$ in terms of 
physical states. Keeping only the $q\bar qg$ part, one has:
\bea
|\Psi_{sing}\ket=\f1{\sqrt{N_c}}\sum_{\alpha\bar\alpha\lambda a}\int d^2z dy\
\f{ig_s}{\pi}
\left[\f{\epsilon_\lambda.(x_0\!-\!z)}{|x_0\!-\!z|^2}
-\f{\epsilon_\lambda.(x_1\!-\!z)}{|x_1\!-\!z|^2}\right]
\left\{\f1{C_FN_c}\mbox{Tr}
\lr{W_F^\dagger(x_1)T^bW_F(x_0)T^c}W_A^{bc}(z)\right.\nonumber\\
\left.-\f1{N_c}\mbox{Tr}\lr{W_F^\dagger(x_1)W_F(x_0)}\right\}
T^a_{\bar\alpha \alpha}
|(\alpha,x_0),(\bar\alpha,x_1),(a,\lambda,z,y)\ket_{phys}\otimes|t\ket\ 
.\label{wrongdiff}\eea

$|\Psi_{sing}\ket$ is not exactly the final state one needs to consider since 
our definition of diffractive implies that the target does not break up and we 
have imposed that condition yet. The final state (\ref{wrongdiff}) would be the 
one to use to calculate diffractive gluon production without any requirement on 
the target. To insure the target not to break up, one projects 
the outgoing state $|\Psi_{sing}\ket$ on the subspace spanned by 
$|t\ket.$ The effect is that the Wilson lines in (\ref{wrongdiff}) become 
target-averaged:
\bea
|\Psi_{diff}\ket=|t\ket\bra t||\Psi_{sing}\ket=
\f1{C_FN_c\sqrt{N_c}}\sum_{\alpha\bar\alpha\lambda a}\int d^2z dy\
\f{ig_s}{\pi}
\left[\f{\epsilon_\lambda.(x_0\!-\!z)}{|x_0\!-\!z|^2}
-\f{\epsilon_\lambda.(x_1\!-\!z)}{|x_1\!-\!z|^2}\right]
\hspace{3cm}\nonumber\\
\times\Phi(z)T^a_{\bar\alpha \alpha}
|(\alpha,x_0),(\bar\alpha,x_1),(a,\lambda,z,y)\ket_{phys}\otimes|t\ket\ \eea
with
\be
\Phi(z)=\left\bra\mbox{Tr}\lr{W_F^\dagger(x_1)T^aW_F(x_0)T^b}
W_A^{ab}(z)\right\ket_t
-C_F\left\bra\mbox{Tr}\lr{W_F^\dagger(x_1)W_F(x_0)}\right\ket_t\ .\label{phi}\ee

The diffractive gluon-production cross-section is now obtained from formula 
(\ref{csec}) with $|\Psi_{diff}\ket$ instead of $|\Psi_{out}\ket.$ 
The final result is
\be
\f{d\sigma^{diff}}{d^2kdy}(x_{01})=\f{\alpha_s}{\pi^2C_FN_c^2}
\int d^2b\int\f{d^2z_1}{2\pi}\f{d^2z_2}{2\pi}\ 
e^{ik.(z_2\!-\!z_1)}
\left[\f{x_0\!-\!z_1}{|x_0\!-\!z_1|^2}-\f{x_1\!-\!z_1}{|x_1\!-\!z_1|^2}\right].
\left[\f{x_0\!-\!z_2}{|x_0\!-\!z_2|^2}-\f{x_1\!-\!z_2}{|x_1\!-\!z_2|^2}\right]
\Phi(z_1)\Phi^*(z_2)\ .\label{finaldiff}\ee

Note that if one had computed the cross-section with $|\Psi_{sing}\ket,$ there 
would have been a global target average at the cross-section level and not 
averages at the amplitude level as is the case in (\ref{finaldiff}).
The first term in (\ref{phi}) is proportional to the elastic ${\cal S}-$matrix 
for the scattering of a colorless $q\bar qg$ triplet on the target while the 
second term is proportional to the elastic ${\cal S}-$matrix for the scattering 
of a $q\bar q$ dipole. There is then two dynamical quantities playing a role in 
this diffractive cross-section. This is the main difference compared to the 
inclusive case in which there was only one. 

If one considers that the target is a nucleus, and that each scattering on the 
nucleons happens via a two-gluon exchange, then the target averages in 
(\ref{phi}) are computable (see {\it e.g.} \cite{kovwie}) and one recovers the 
result of \cite{kovch}. Formulae (\ref{finaldiff}) and (\ref{phi}) are a 
generalization to an arbitrary target including all numbers of gluon exchanges.

Let us mention also that, making use of the following identity:
\be
2\mbox{Tr}\lr{W_F^\dagger(x_1)T^aW_F(x_0)T^b}
W_A^{ab}(z)=\mbox{Tr}\lr{W_F^\dagger(x_1)W_F(z)}
\mbox{Tr}\lr{W_F^\dagger(z)W_F(x_0)}
-\f1{N_c}\mbox{Tr}\lr{W_F^\dagger(x_1)W_F(x_0)}\ ,\ee
one is able to relate the first term of (\ref{phi}) to the scattering of two 
dipoles and obtain:
\be
\Phi(z)=\f12\left\bra\mbox{Tr}\lr{W_F^\dagger(x_1)W_F(z)}
\mbox{Tr}\lr{W_F^\dagger(z)W_F(x_0)}\right\ket_t
-\f{N_c}2\left\bra\mbox{Tr}\lr{W_F^\dagger(x_1)W_F(x_0)}\right\ket_t\ .\ee
The first term in $\Phi$ is now proportional to the elastic ${\cal S}-$matrix 
for the scattering of two $q\bar q$ dipoles on the target. What is usualy 
considered is the scattering off a large nucleus in which case one can justify 
neglecting the fluctuations and writing
\be
\left\bra\mbox{Tr}\lr{W_F^\dagger(x_1)W_F(z)}
\mbox{Tr}\lr{W_F^\dagger(z)W_F(x_0)}\right\ket_t=
\left\bra\mbox{Tr}\lr{W_F^\dagger(x_1)W_F(z)}\right\ket_t
\left\bra\mbox{Tr}\lr{W_F^\dagger(z)W_F(x_0)}\right\ket_t\ .\label{nocor}
\ee
Making this simplification is very useful to do phenomenology \cite{munsho}, as 
it enables to deal with only one quantity, but it is not correct for any target.
Note also that if one writes the product in the right-hand side of (\ref{nocor}) 
using $T$ amplitudes as in (\ref{gdip}), one recovers the two-gluon exchange 
approximation calculated in \cite{bart,kopdiff} by neglecting the $T^2$ term.

\section{Dipole-factorized form of the inclusive cross-section}
\label{4}

We now come back to the case of the inclusive cross-section (\ref{final}) and 
show in this section that it can be significantly simplified and expressed in a 
form that exhibits a dipole factorization.

On the right-hand side of (\ref{final}), the term 
in brakets contains four amplitudes and for each of them two of the integrations 
can easily be carried out. Let us begin with the amplitude which does not 
depend on $b,$ the integration of the emission kernel over that impact parameter 
gives:
\be
\int d^2b\ \f{(x_i\!-\!z_1).(x_j\!-\!z_2)}{|x_i\!-\!z_1|^2|x_j\!-\!z_2|^2}=
2\pi\log{\f{\rho}{|x_{ij}\!-\!z_1\!+\!z_2|}}\ee
where $x_{ij}\!=\!x_i\!-\!x_j$ and $\rho$ is an infrared cutoff which will 
eventually disappear. Then one writes:
\be
\int\f{d^2z_1}{2\pi}\f{d^2z_2}{2\pi}\ e^{ik.(z_2\!-\!z_1)}
{2\pi}\log{\f{\rho}{|x_{ij}\!-\!z_1\!+\!z_2|}}
T(z_1,z_2,Y\!-\!y)=\f12
\int\f{d^2z}{2\pi}\ e^{-ik.z}\log{\f{\rho}{|x_{ij}\!-\!z|}}\ 
\sigma_{(gg)t}(z,Y\!-\!y)
\ .\ee
Consider now the amplitude which only depends on $z_1,$ we then perform 
the $z_2$ integration:
\be \int\f{d^2z_2}{2\pi}\ e^{ik.(z_2\!-\!z_1)}
\f{(x_i\!-\!z_1).(x_j\!-\!z_2)}{|x_i\!-\!z_1|^2|x_j\!-\!z_2|^2}
=\f{k.(x_i\!-\!z_1)}{ik^2|x_i\!-\!z_1|^2}\ e^{ik.(x_j-z_1)} \ee
and obtain
\be 
\int d^2b\int\f{d^2z_1}{2\pi}
\f{k.(x_i\!-\!z_1)}{ik^2|x_i\!-\!z_1|^2}\ e^{ik.(x_j-z_1)}
T(z_1,x_j,Y\!-\!y)=\f12
\int\f{d^2z}{2\pi}\ e^{-ik.z}\f{k.(x_{ij}\!-\!z)}{ik^2|x_{ij}\!-\!z|^2}
\ \sigma_{(gg)t}(z,Y\!-\!y)\ .
\ee
One can easily see that the amplitude which only depend on $z_2$ gives 
the same result. For the last amplitude which does not depend either on 
$z_1$ or $z_2,$ we integrate the emission kernel:
\be
\int\f{d^2z_1}{2\pi}\f{d^2z_2}{2\pi}\ e^{ik.(z_2\!-\!z_1)}
\f{(x_i\!-\!z_1).(x_j\!-\!z_2)}{|x_i\!-\!z_1|^2|x_j\!-\!z_2|^2}=
\f1{k^2}\ e^{-ik.x_{ij}}
\ee
and the $b$ integration of the amplitude yields to the cross-section.
Putting the pieces together, the gluon production cross-section reads
\be
\f{d\sigma}{d^2kdy}(x_{01})=\f{\alpha_sC_F}{2\pi^2}\int\f{d^2z}{2\pi}\ 
e^{-ik.z}\sum_{i,j=0}^{1}(-1)^{i+j}
\lr{\log{\f{|x_{ij}\!-\!z|}{\rho}}
-\f{2ik.(x_{ij}\!-\!z)}{k^2|x_{ij}\!-\!z|^2}
-\f{2\pi}{k^2}\ \delta^{(2)}(x_{ij}\!-\!z)}\sigma_{(gg)t}(z,Y\!-\!y)\ .
\ee
Performing the summation over $i$ and $j,$ the cutoff $\rho$ disappears 
and the two terms displaying $\delta^{(2)}(z)$ also vanish because 
$\sigma_{(gg)t}(0,y)\!=\!0$. Using 
then the following results for two dimensionnal vectors where $\Theta$ 
is the Heavyside step function:
\be 
\int_0^{2\pi}d\theta_x\log{\f{|z|^2}{|x\!-\!z||x\!+\!z|}}=
4\pi\Theta(|x|\!-\!|z|)
\log{\f{|z|}{|x|}}\ ,\ee
\be \int_0^{2\pi}d\theta_x\ \f{2i}{k^2}\lr{2\f{k.z}{|z|^2}
-\f{k.(z\!-\!x)}{|z\!-\!x|^2}-\f{k.(z\!+\!x)}{|z\!+\!x|^2}}=\f{8i\pi}{|k||z|}
\cos{\theta_z}\Theta(|x|\!-\!|z|)\ ,\ee
one carries out the integration over the azimutal angle of $x_{01}.$ 
Denoting the size of the incident $q\bar q$ dipole $|x_{01}|\!\equiv\!r_0,$ the 
result reads
\be
\f{d\sigma}{d^2kdy}(r_0)=\f{\alpha_sC_F}{\pi^2}\int d^2z\ 
e^{-i|k||z|\cos{\theta}}
\left\{\lr{\f{2i}{|k||z|}\cos{\theta}-\log{\f{r_0}{|z|}}}\Theta(r_0\!-\!|z|)
+\f2{k^2}\delta(z^2\!-\!r_0^2)\right\}
\sigma_{(gg)t}(z,Y\!-\!y)\ .
\ee
Most of the time the cross-section $\sigma_{(gg)t}(z,y)$ will not depend on 
$\theta$ as one averages over the azimutal angle of the target. This enables  
the final $\theta$ integration: 
\be 
\f{d\sigma}{d^2kdy}(r_0)=\f{2\alpha_sC_F}{\pi k^2}
\int dz^2\phi(r_0,z,k)\sigma_{(gg)t}(z,Y\!-\!y)\ .\label{dipfact}\ee
We have singled out the following normalized distribution: 
\be
\phi(r_0,z,k)=\Theta(r_0\!-\!z)\lr{\f kz J_1(kz)-\f{k^2}2 
J_0(kz)\log(r_0/z)}+J_0(kz)\delta(r_0^2-z^2)\ ,\label{dipdis}\ee
\be
\int dz^2\ \phi(r_0,z,k)=1\ .\ee
Note that now, $z$ and $k$ respectively stand for $|z|$ and $|k|.$ The 
gluon production cross-section as written in (\ref{dipfact}) exhibits a dipole
factorization: the $gg$ dipole-target cross-section $\sigma_{(gg)t}$ is 
convoluted with the effective dipole distribution $\phi$ which contains the 
emission of the forward jet of momentum $k$ off the initial $q\bar q$ dipole of 
size $r_0$. It is remarkable that we obtain factorization and such a simple 
formula.

One can easily check
that $\phi$ (\ref{dipdis}) verifies:
\be
2z\phi(r_0,z,k)=\f{\partial}{\partial z}
\left[\Theta(r_0-z)z\f{\partial}{\partial z}
\lr{J_0(kz)\log{\f{r_0}z}}\right]\ .\ee
Thanks to this identity, one can express the gluon production cross-section 
(\ref{dipfact}) in another convenient way:
\be 
\f{d\sigma}{d^2kdy}(r_0)=\f{2\alpha_sC_F}{\pi k^2}
\int_0^{r_0} dz\ J_0(kz)\log{\f{r_0}z}\ \f{\partial}{\partial z}\lr{z
\f{\partial}{\partial z}\ \sigma_{(gg)t}(z,Y\!-\!y)}\ .\label{simp}\ee
Formulae (\ref{dipfact}) and (\ref{simp}) express the single inclusive 
forward-gluon production cross-section in the scattering of a $q\bar q$ dipole 
off an arbitrary target and are the main results of the paper. The intermediate 
cross-section $\sigma_{(gg)t}$ for the scattering of a $gg$ dipole on 
the target includes all multiple scatterings and linear or non-linear 
quantum evolution. As shown by (\ref{gdip}) and (\ref{csection}), it can be 
computed from a trace of Wilson lines correctly averaged over the target gluon
fields.

So far, we have only considered the case where the measured gluon is the closest
in rapidity to the dijet coming from the dissociation of the incident $q\bar q$ 
pair. One can extend the result to the case of a gluon measured at a less 
forward rapidity by including the emissions of harder gluons, emitted before the 
gluon detected in the final state. To stay consistent with our soft-gluon 
approximation, this can be done by including leading logarithmic evolution 
before the emission of the measured gluon. This is quite straightforward if one 
sticks to linear BFKL evolution which is justified as long as the rapidity of 
the measured gluon is not to big and non-linearities are negligible. Since the 
incident particule is a $q\bar q$ dipole, this BFKL evolution is the usual 
dipole evolution from the initial dipole of size $r_0$ to the dipole of size $r$ 
from which the measured gluon is now emitted. In formulae, that amounts to do 
the substitution:
\be
\f{d\sigma}{d^2kdy}(r_0)\rightarrow\int\f{d^2r}{2\pi r^2}\ 
n(r_0,r,y)\f{d\sigma}{d^2kdy}(r)\label{subs}\ee
where $n(r_0,r,y)$ is the number density of dipoles of size $r$ inside the 
initial dipole of size $r_0$ at rapidity $y.$ This quantity was defined in 
\cite{mueller} and was shown to satisfy the BFKL equation \cite{bfkl}. The 
solution is:
\be
n(r_0,r,y)=2\intc{\g}\lr{\f{r_0}r}^{2\g} 
\exp\lr{\f{2\alpha_sC_F}{\pi}y\chi(\g)}\ee
where the complex integral runs along the imaginary axis from $1/2\!-\!i\infty$ 
to $1/2\!+\!i\infty$ and where the BFKL kernel is 
$\chi(\g)=2\psi(1)\!-\!\psi(\g)\!-\!\psi(1\!-\!\g).$ The substitution 
(\ref{subs}) gives:
\be 
\f{d\sigma}{d^2kdy}(r_0)=\f{2\alpha_sC_F}{\pi k^2}
\int dz\ J_0(kz)\tilde{n}(r_0,z,y) \f{\partial}{\partial z}\lr{z
\f{\partial}{\partial z}\ \sigma_{(gg)t}(z,Y\!-\!y)}\label{evol}\ee
where 
\be
\tilde{n}(r_0,z,y)=\intc{\g}\lr{\f{r_0}z}^{2\g}\f1{2\g^2}
\exp\lr{\f{2\alpha_sC_F}{\pi}y\chi(\g)}\ .\ee

The most natural application to our formulae is a measurement in deep 
inelastic scattering in which one detects the most closest jet to the dijet 
coming from the dissociation of the photon in case of formula (\ref{simp}) or a 
less forward jet for formula (\ref{evol}). Such a measurements would certainly 
give information on the dipole-proton scattering. However the rest of our study 
will not concentrate on that, it will turn to the case of forward jets emitted 
by hadrons. Such jets represent the most natural probes in hadronic collisions 
and are a very interesting application to our formulae.

\section{An application: forward-jet production}
\label{5}

In this part we consider the same cross-section as in the previous section, 
but with the initial particule being a hadron. Indeed in experiments,
collisions that procuce jets are proton-lepton or proton-antiproton
collisions. Forward jets are jets measured in the forward directions of the
hadrons. We cannot compute such processes directly because of the 
non-perturbative nature of the hadrons. Therefore we start by explaining how to 
extend the previous result from an incident perturbative $q\bar q$ dipole to an 
incoming hadron. In a second part, we also specify the target and
consider a virtual photon, the goal being to describe forward jets in deep
inelastic scaterring. We finally generalize to Mueller-Navelet jets at hadron 
colliders.

\subsection{Gluon production off a hadron}

In this section we explain how to make the link between the $q\bar q$ dipole and 
the hadron using collinear factorization. Indeed, providing that the transverse 
momentum of the measured jet is larger than a perturbative scale, forward-jet 
production is a hard process and obeys collinear factorization. Moreover, since 
we are dealing with forward jets, the final state gluon has a fraction of 
longitudinal momentum (with respect to the incident hadron) $x_J$ that is never 
too low, thus there are no small$-x$ effects in the wavefunction of the hadron 
that could break collinear factorization.

Here is how we proceed. Consider first the cross-section (\ref{evol}) derived in 
the previous section and take the collinear 
limit for the incident dipole $r_0\mu\!\gg\!1$ where $\mu$ is an arbitrary hard 
scale called the factorization scale. This limit requires the emission of the 
gluon to occur at a scale $k\!\sim\!\mu$ much larger than the scale of the 
initial dipole $1/r_0.$ Such a strong ordering between the two scales is indeed 
what leads to the collinear factorization. Moreover it is justified in this case 
because we really want to describe the emission of the jet off a hadron which 
naturally carries a soft scale. In this appropriate limit, $\tilde{n}(r_0,z,y)$ 
reduces to:
\be
\tilde{n}(r_0,z,y)=\intc{\g}\f{(r_0\mu)^{2\g}}{2\g^2}\
\exp\lr{\f{2\alpha_sC_F}{\pi}y\chi(\g)}(z\mu)^{-2\g}\simeq
\intc{\g}\lr{r_0\mu}^{2\g}\f1{2\g^2}
\exp\lr{\f{2\alpha_sC_F}{\pi\g}y}\label{ione}\ee
since for $r_0\mu\!\gg\!1$ the integral is determined by the $\g\!\simeq\!0$ 
behavior of the integrand. We have used $\chi(\g)\!\simeq\!1/\g$ for 
$\g\!\simeq\!0.$ Then one recognizes that the right-hand side of (\ref{ione}) is 
an integral representation of a modified Bessel function \cite{bessel}:
\be
\intc{\g}\lr{r_0\mu}^{2\g}\f1{2\g^2}
\exp\lr{\f{2\alpha_sC_F}{\pi\g}y}=\f12
\sqrt{\f{\pi\log(r_0^2\mu^2)}{2\alpha_sC_Fy}}\
I_1\lr{2\sqrt{\f{2\alpha_sC_F}{\pi}\log(r_0^2\mu^2)y}}.\ee
And (\ref{evol}) reduces finally to:
\bea
\f{d\sigma}{d^2kdx_J}(r_0)\simeq\f1{2k^2x_J}
\left\{
\sqrt{\f{2\alpha_sC_F\log(r_0^2\mu^2)}{\pi\log(1/x_J)}}\
I_1\lr{2\sqrt{\f{2\alpha_sC_F}{\pi}\log(r_0^2\mu^2)\log(1/x_J)}}\right\}
\hspace{2cm}\nonumber\\
\times\int dz\ J_0(kz)\ \f{\partial}{\partial z}\lr{z
\f{\partial}{\partial z}\ \sigma_{(gg)t}(z,\Delta\eta)}\label{colfact}\eea
where we have changed the variable $y\!=\!\log{1/x_J}$ and defined
$\Delta\eta\!=\!Y\!-\!\log{1/x_J},$ the rapidity interval between the jet and 
the target. The factor in brackets in (\ref{colfact}) is the gluon distribution 
function $x_Jg_d(x_J,\mu^2)$ inside the incident $q\bar q$ dipole of size $r_0$ 
at the Double Leading Logarithmic (DLL) approximation. This is not surprising: 
we have assumed that the emitted gluon was soft which is responsible for the 
leading $\log(1/x_J)$ and we have considered the collinear limit which is 
responsible for the leading $\log(r_0^2\mu^2).$ Within this DLL approximation, 
the distribution $g_d$ inside the $q\bar q$ pair of size $r_0$ is the same than 
the gluon distribution inside a gluon of virtuality $1/r_0^2$ with $2\ C_F$ 
being the color factor instead of $C_A.$ Now the link with the gluon production 
cross-section off an incident hadron is obvious: the hard part of the 
cross-section is the same and the hadron appears in the cross-section via its 
parton distribution function $g_h$. One has then
\be
\f{d\sigma^{h\!+\!t\!\rightarrow\!J\!+\!X}}{d^2kdx_J}=\f1{2k^2}\ 
g_h(x_J,\mu^2)
\int dz\ J_0(kz)\ \f{\partial}{\partial z}\lr{z
\f{\partial}{\partial z}\ \sigma_{(gg)t}(z,\Delta\eta)}\ .\label{fjet}\ee
Note that this formula is only valid at the DLL approximation since we have 
always taken the emitted gluon to be soft, that is $x_J\!\ll\!1.$

A simple way to see that (\ref{fjet}) is the correct formula is to 
consider the case of linear leading-logarithmic BFKL evolution between the jet 
and the target. 
In this case $\sigma_{(gg)t}$ obeys $k_T-$factorization and can be 
expressed 
in terms of the unintegrated distribution in the target $f_T:$
\be
\sigma_{(gg)t}^{BFKL}(z,\Delta\eta)=\f{4\alpha_SN_c}{\pi}
\int\f{d^2q}{q^4}(1-J_0(qz))f_T(q,Q_T,\Delta\eta)
\label{diptar}\ee
where $Q_T$ is a scale characterising the target. Inserting (\ref{diptar})
into (\ref{fjet}) gives
\be
\f{d\sigma^{h\!+\!t\!\rightarrow\!J\!+\!X}_{BFKL}}{d^2kdx_J}=
\f{4\alpha_SN_c}{k^4}\ g_h(x_J,\mu^2)f_T(k,Q_T,\Delta\eta)
\ee
and we recover the known BKFL result \cite{theo} for forward-jet production.

Formula (\ref{fjet}) can be written in a very simple factorized form if one 
considers the cross-section integrated over the transverse momentum of the jet 
above a given cut $E_T:$
\be
\f{d\sigma^{h\!+\!t\!\rightarrow\!J\!+\!X}}{dx_J}\equiv
\int_{E_T^2}^\infty dk^2\f{d\sigma^{h\!+\!T\!\rightarrow\!J\!+\!X}}{d^2kdx_J}= 
g_h(x_J,\mu^2)
\int d^2z\ \phi_J(z,E_T)\sigma_{(gg)t}(z,\Delta\eta)\ ,\label{fjint}\ee
where the distribution $\phi_J$ is related to a bessel function:
\be
\phi_J(z,E_T)=\f{E_T}{2\pi z}J_1(E_Tz).\ee
We have assumed that the $gg$ dipole-target cross-section 
$\sigma_{(gg)t}(z)$ 
goes to a constant as $z$ goes to infinity fast enough to cancel the edge terms 
in the integrations by parts. Typically, it works with 
$\sigma_{(gg)t}(z)\!\sim\!1\!-\!\exp{(-z^2)}$ or with 
$\sigma_{(gg)t}(z)\!\sim\!1\!-\!\exp{(-\log^2(z))}.$ Formula (\ref{fjint})
exhibits the dipole factorization for a forward jet emitted off a hadron. We 
recover the result obtained from indirect calculations in \cite{munpes} using 
the equivalence between the $k_T-$factorization and the dipole factorization. 
The feature that this approach couldn't reveal in is that it is a $gg$ dipole 
that has to be considered and not a $q\bar q$ dipole. We have also shown here 
that the dipole factorization is still valid when the cross-section 
$\sigma_{(gg)t}$ is not $k_T-$factorizable, justifying the use of (\ref{fjint}) 
to study saturation effects in forward jets \cite{marpes,mpr}.

\subsection{Forward jets in deep inelastic scattering}

The study of forward jets in deep inelastic scattering represents the most 
straightforward application of our formula. In such processes, a proton scatters 
on an electron and a jet is detected in the forward direction of the proton. 
Therefore the interaction takes place between 
the proton and a transversally ($T$) or longitudinally ($L$) polarized virtual 
photon, see Fig.3a. An hadronic interaction with a virtual photon is easily 
expressed in terms of $q\bar q$ dipoles: the photon fluctuates into the 
quark-antiquark pair which then interacts. The wave functions $\phi_T$ and 
$\phi_L$ describing the splitting are well-known. The cross-section is then 
given by formula (\ref{fjint}) with the $gg$ dipole-$q\bar q$ dipole 
cross-section $\sigma_{(gg)d}$ carrying the rapidity dependance and with an 
extra convolution with $\phi_{T,L}:$
\be
\f{d\sigma^{p\!+\!\g^*\!\rightarrow\!J\!+\!X}_{T,L}}{dx_J}=
g_p(x_J,\mu^2)
\int d^2rd^2z\ 
\phi_{T,L}(r,Q^2)\phi_J(z,E_T)\sigma_{(gg)d}(r,z,\Delta\eta)\ 
.\label{fjetdis}\ee
The virtuality of the photon has been denoted $Q^2$ and $E_T$ is the cut on the 
jet transverse momentum. In the regime where $Q\!\sim\!E_T,$ DGLAP 
evolution is suppressed and fixed-scale quantum evolution drives the high-energy 
behavior of the cross-section. This evolution is contained in 
$\sigma_{(gg)d}$ which is the only unknown quantity in formula 
(\ref{fjetdis}). It can be obtained by computing a trace of Wilson lines 
averaged over the fields of the dipole, see formulae (\ref{gdip}) and 
(\ref{csection}); performing the target averages and calculating 
$\sigma_{(gg)d}$ is however not easy. The color glass condensate theory 
provides a way of doing it:
\be
\left\bra\mbox{Tr}\lr{W_A^\dagger(x')W_A(x)}\right\ket_d(\Delta\eta)=
\int{\cal D}[{\cal A}]U_{\Delta\eta}[{\cal A}]\ 
\mbox{Tr}\lr{W_A^\dagger[{\cal A}(x')]W_A[{\cal A}(x)]}
\label{cgc}\ee
where $U_{\Delta\eta}[{\cal A}]$ is a functional which specifies the 
probability to have a given field configuration in the target $q\bar q$ dipole. 
When $\Delta\eta\simeq 0$, there are no extra gluon radiation and the amplitude 
provides the initial condition to the JIMWLK equation \cite{jimwlk} that 
describes how $U_{\Delta\eta}[{\cal A}],$ and thus also the target average 
(\ref{cgc}), varies as one increases the rapidity interval $\Delta\eta.$ A lot 
of work is employed trying to solve the JIMWLK equation, at least numerically 
but this is far beyond the scope of this paper. Providing the rapidity interval 
$\Delta\eta$ is not too high, the JIMWLK equation reduces to the BFKL equation 
whose solution has been known for years. A first solution valid beyond BFKL has 
been found in \cite{edal} but it is only valid as long as the color fields in 
the two dipoles are weak.

\begin{figure}[ht]
\begin{center}
\epsfig{file=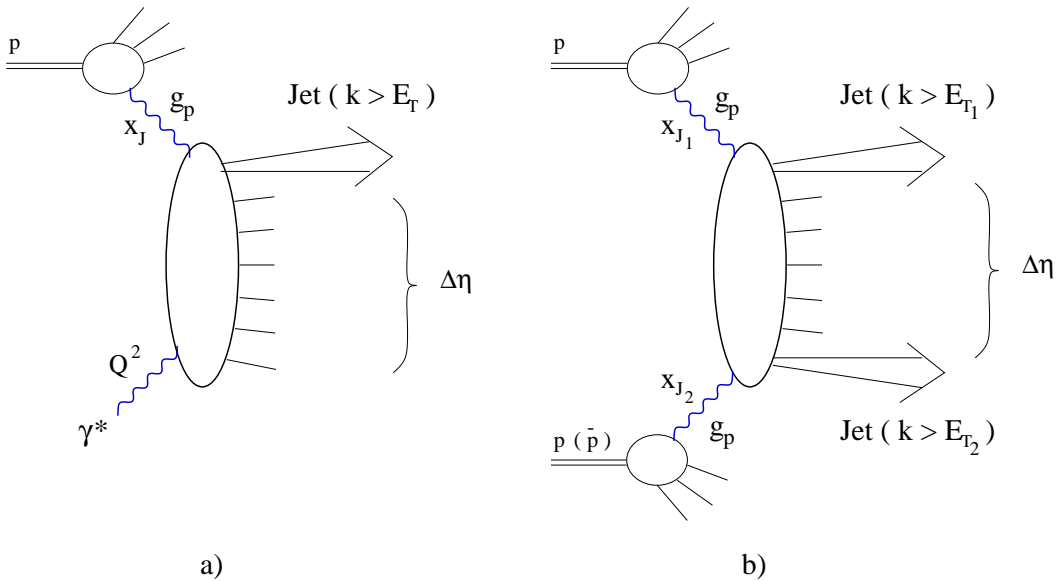,width=14cm}
\caption{a) Forward-jet production in a 
proton-lepton collison (at HERA). b) Mueller-Navelet jet production in a 
proton-proton (at LHC) or proton-antiproton (at Tevatron) collision.
$E_T$, $E_{T_1}$, and $E_{T_2}$: cuts on the transverse 
momenta of the jets; $x_J,$ $x_{J_1},$ and $x_{J_2}$: longitudinal momentum 
fraction of the jets with respect to the incident hadron; $\Delta\eta:$ rapidity 
intervals between the hard probes.}
\end{center}
\label{F3}
\end{figure}

Anyway, as shown by formula (\ref{fjetdis}), the forward-jet measurement is a 
suitable observable to study the $gg$ dipole-$q\bar q$ dipole scattering and 
saturation effects purely due to quantum evolution. Forward-jets at HERA seem to 
be the good candidate to perform the present studies \cite{mpr}.

\subsection{Mueller-Navelet jets at hadron colliders}

In the future, Mueller-Navelet jets \cite{mnj} at LHC could be decisive while at 
the Tevatron the rapidity range is probably not large enough \cite{marpes,mpr}. 
Mueller-Navelet jets are processes in which a proton strongly interacts with 
another proton or antiproton and where a jet with a hard transverse momemtum is 
detected in each of the two forward directions, see Fig.3b. 
Such processes can also be computed from formula (\ref{fjet}): the target is 
itself a hadron emitting a forward jet. The two cuts on the transverse momenta 
of the jets will be denoted $E_{T_1}$ and $E_{T_2}$ and $\Delta\eta$ is now the 
rapidity interval between the two jets. One has
\be
\f{d\sigma^{p\!+\!p\!\rightarrow\!J\!+\!X+\!J}}{dx_{J_1}dx_{J_2}}=
g_p(x_{J_1},\mu^2)g_p(x_{J_2},\mu^2)
\int d^2rd^2z\ \phi_J(r,E_{T_1})\phi_J(z,E_{T_2})
 \sigma_{(gg)(gg)}(r,z,\Delta\eta)\label{mnjets}\ee
and now the cross-section $\sigma_{(gg)(gg)}$ for the scattering of 
two $gg$ dipoles is the quantity driving the evolution with the rapidity. 
(\ref{fjetdis}) and (\ref{mnjets}) are generalizations of the known formulae for 
forward jets \cite{dis} and Mueller-Navelet jets \cite{mnj} that included only 
linear BFKL evolution. Full quantum evolution is included in (\ref{fjetdis}) and 
(\ref{mnjets}) via the $gg$ dipole-$gg$ dipole cross-section. That makes those 
two measurements suitable observables for studying saturation due to evolution 
and unitarization of the dipole-dipole scattering, as well as for testing 
phenomenological models \cite{marpes}.

\section{Conclusions}
\label{6}

We have derived inclusive and diffractive forward-gluon production in the 
scattering of a $q\bar q$ dipole off a unspecified target in the high-energy 
eikonal approximation. We worked in a framework in which the incident $q\bar q$ 
pair is represented by its light-cone wavefunction and in which the interaction 
with the target is described by Wilson lines that shift the components of the 
incident wavefunction with eikonal phases. We have also assumed that the emitted 
gluon was soft so that our results are valid at leading-logarithmic accuracy for 
this measured gluon. The results are given by formulae (\ref{final}) and 
(\ref{finaldiff}) for the inclusive and diffractive cross-sections respectively. 
All multiple gluon exchanges with the target are included. The measured gluon 
being always the most-forward gluon with respect to the incident $q\bar q$ 
dipole, quantum evolution is contained in the target-averaged quantities. For 
the inclusive cross-section, only one such quantity is involved: the forward 
scattering amplitude of a colorless pair of gluons on the target (\ref{gdip}). 
In the diffractive case however, more complicated $q\bar qg$ amplitudes play 
also a role (\ref{phi}).

In the inclusive case, without specifying the target, we were able to simplify 
the result (\ref{final}) and to write the cross-section in the factorized form 
(\ref{dipfact}). It expresses the forward-jet cross-section as the convolution 
between an effective dipole distribution (\ref{dipdis}) and the $gg$
dipole-target cross-section. The dipole factorization (\ref{dipfact}) for a 
forward-jet emission is the equivalent of the usual dipole factorization for a 
virtual photon: this is a dipole formalism for a hadronic probe. 

As a simple application of this result, we described forward jets at hadronic 
colliders. We extended the result to the case of an incident hadron by including 
linear evolution before the emission of the measured gluon jet and then by using 
the collinear factorization properties of QCD. Formula 
(\ref{fjint}) is thus is very simple dipole-factorized cross-section that 
describes forward jets at HERA, Tevatron or LHC, depending on the target. We 
finally gave the expression for those forward-jet cross-section (\ref{fjetdis}) 
and Mueller-Navelet jet cross-section (\ref{mnjets}). They are written in terms 
of the $gg$ dipole-$q\bar q$ dipole or $gg$ dipole-$gg$ dipole cross-sections 
which are the quantities that determine the high-energy behavior via their 
rapidity evolution. We have therefore established a link between two observables 
measurable at HERA, Tevatron or LHC and dipole-dipole cross-sections which are 
the basic quantities to study to understand high-energy scattering in QCD, 
saturation and unitarization due to quantum evolution.

\begin{acknowledgments}

A part of this work was carried out during the Theory Summer Program on RHIC 
Physics and I wish to thank the theory group at BNL for their hospitality during 
that worshop. I specially thank Robi Peschanski for many useful comments and 
suggestions and St\'ephane Munier for bringing to my attention the work of 
references \cite{kovch,kovtuch}. I also gratefully acknowledge informative and 
helpful discussions with Kirill Tuchin, Larry McLerran, Genya Levin, Boris 
Kopeliovich, Fran\c{c}ois G\'elis, Raju Venugopalan, and Edmond Iancu.
 
\end{acknowledgments}

\end{document}